\documentclass{ws-procs9x6}

\usepackage{epsfig}
\usepackage{amssymb}
\usepackage{amsmath}
\usepackage{graphicx}
\usepackage{amsthm}
\usepackage{epsfig}

\title{\bf SHORT AND LONG-TERM DYNAMICS OF CHILDHOOD DISEASES IN DYNAMIC SMALL-WORLD NETWORKS}

\author{ J. VERDASCA} 

\address{Centro de Matem\'{a}tica e Aplica\c{c}\~{o}es Fundamentais \\  Complexo Interdisciplinar da Universidade de Lisboa \\  Av. Professor Gama Pinto 2, P-1649-003 Lisboa, Portugal\footnote{\uppercase{P}resent address: \uppercase{C}entro de \uppercase{A}strobiologia, \uppercase{I}nstituto Nacional de \uppercase{T}\'{e}cnica \uppercase{A}eroespacial, \uppercase{C}tra. de \uppercase{T}orrej\'{o}n a \uppercase{A}jalvir, km 4, 28850 \uppercase{T}orrej\'{o}n de \uppercase{A}rdoz, \uppercase{M}adrid, \uppercase{S}pain.}}

\begin{document}

\maketitle

\abstracts{We have performed  individual-based  lattice simulations of SIR and SEIR dynamics to investigate both the
short and long-term dynamics of childhood epidemics. In our model, infection takes place through  a  combination of local  and long-range contacts, in practice
generating  a dynamic  small-world network. Sustained oscillations emerge  with a period much larger than the duration of infection. We found that the network topology has a strong   impact on the amplitude of oscillations  and in  the level of persistence. 
Diseases do not  spread very effectively through local contacts. This can be seen by measuring  an  {\em effective} transmission rate $\beta_{\mbox {\scriptsize eff}}$ as well as  the basic reproductive rate $R_0$. These quantities are lower in the small-world network   than in an homogeneously mixed population, whereas the average age at infection is higher. \textbf{Keywords}: Lattice model; Recurrent epidemics; Small-world networks 
}

\section{Introduction}

The recurrent outbreaks of measles and other childhood diseases is one of the most striking features of the pre-vaccination records. Despite  continuing efforts over more than seven decades,  the question of what is the   mechanism behind these  oscillations   has not yet received a fully satisfactory answer \cite{AM}.

Measles, mumps, varicella (chickenpox) and  rubella are examples of  diseases that confer lifelong immunity, and can be analyzed within the  SIR (Susceptible, Infected and Recovered)  or SEIR (comprising an additional class, the Exposed) general frameworks. If the population is  constant, the  mean-field implementation of   SIR and SEIR scheme disregarding heterogeneity in contact structure
leads to  systems of respectively two and three coupled ODE's. 
 Although based on this most unrealistic description  of human contacts, mean-field deterministic  models still capture most {\em static} properties of epidemics of typical childhood diseases including the  threshold values for the spread of an epidemic, its final size 
as well as  the average age at which infection is acquired. However, these models fail in that they predict  oscillations that are invariably damped. This, of course,  contradicts the  available records which evidence self-sustained oscillations of roughly  constant period during the pre-vaccination era.

 After a period during which complicated age-structured models were favored,
seasonally forced models have become the framework of choice to explain  the historic time-series. In fact, when subject to parametric  forcing, deterministic SIR and SEIR ODE's display a rich dynamical behaviour including period-doubling cascades to chaos, quasiperiodicity, multistability between cycles of different periods, etc. \cite{ERBG,GKB}.  When the drive is sinusoidal,  the level of forcing at which  complex behaviour is  observed is deemed unreasonable. However, when a more realistic formulation is used, like an  alternating sequence of periods of high and low transmissibility  mimicking the opening and closing of schools,  the levels of forcing required to obtain  complex behaviour is   considerably lowered \cite{ERBG,KRG}. If epidemics  correspond to  periodic orbits perturbed by noise, as opposed to chaotic solutions \cite{OlsenShaff,RohKeelGren}, then  only  periods which are integer multiples of the forcing period are allowed.  More likely, the observation of both integer and non-integer periods in incidence time series  of rubella and chickenpox  is  the signature  of an  autonomous system oscillating  with a frequency that may or may not  become locked with an  external drive \cite{Blasius}.

By definition, there is one  important feature that cannot be described by deterministic models. That is the pattern of disease persistence, {\em i.e.} the probability that a recurrent  epidemic goes extinct after a given  number of  cycles. Persistence is an emergent property arising from the interaction of  stochastic effects and dynamics. Its meaningfulness derives from being  the key ingredient in the definition of  the Critical Community Size (CCS), the population number below which a particular disease cannot be sustained. But persistence is also  a tool to assess the relative merits of stochastic  versus deterministic models.  Indeed, by far  the most serious shortcoming of  deterministic forcing is that  during the minima often the fraction of infective individuals falls below  $10^{-10}$,  meaning that the global human population would not be enough to sustain recurrent epidemics.

It is well known that, like forcing,  stochastic effects  also tend to sustain the oscillations \cite{Bailey}. However, the implementation of stochastic  SIR and SEIR dynamics disregarding heterogeneity in  contact structure,  generates only fluctuations which are much too small and irregular when compared to real epidemics unless   immigration of infectives from outside is introduced.

Realizing that deterministic forced models fail short of explaining the  patterns of  persistence and  that stochasticity alone could not provide for the necessary agreement with data, researchers  began to explore the role of spatial factors in the persistence and dynamics  of epidemics. The traditional way to account for  an explicit spatial dependence  in population ecology and epidemiology is through  the use of  metapopulation,  or patch, models. These are based on a coarse-grained distribution of the global population over a number of  interacting subpopulations -- the patches. Within-patch dynamics is built into the model {\em a priory}  and  can be made as complex as one wishes to  from the start, with heterogeneity effects  restricted  to the coupling between the patches.
The coarse-graining procedure limits the  ability of the model to assess the impact of   the structure of individual contacts on the overall dynamics. It says very little  about how   emergent complex behaviour on a global scale can  arise  from simple interaction rules, either strictly local or not. To address this issue  one must  consider instead  a network  of interacting individuals.

Nevertheless, some interesting results have been achieved  by adding  metapopulation structure to stochastic models, either in conjunction with external forcing or not.  Lloyd and May \cite{LloydMay} simulated a two-patch stochastic SEIR model, each one having at least  $10^6$ nodes. The oscillations they  obtained  were  too irregular and their amplitude too small when compared with data records.Moreover, strong coherence  could not  be   obtained  unless a considerable  level of seasonal forcing was applied.  However, in that case the number of infective individuals dropped to the unrealistic levels predicted  by   deterministic  models. Bolker and Grenfell \cite{BolkGren}  considered a similar model but allowed for contacts inside each subpopulation and between individuals belonging to different subpopulations. By increasing the   ratio of between- and within-patch contacts they could increase the levels of  persistence. Although their study clearly indicated  that adding   structure to the network of contacts could indeed enhance persistence it  greatly overestimated the size of the population needed to sustain recurrent  epidemics.

It is interesting to note that the modelling of disease spread  as a combination of local and long-range interactions in the context of  patch models\cite{BolkGren}  actually precedes the introduction  by Watts and Strogatz of a class of networks  that interpolates between regular lattices and random networks \cite{WS} --  small-world (SW) networks  --  and the acknowledgment of the importance of the small-world phenomenon  on the spread of  epidemics that soon  followed \cite{Watts,Moore,Newman,BS,KleGren}. Subsequently {Boots \& Sasaki} \cite{BS} have used  SW networks to analyze  the selection of a particular strain of a pathogen and Keeling \cite{Keeling99} considered  a network of nodes which had many of  the properties of a SW to calculate epidemic thresholds and  determine a number of  properties of the endemic state.

More recently, probabilistic cellular automata (PCA) models of infectious  diseases    evolving  on SW networks were proposed by Johansen \cite{Johan1,Johan2},  Kuperman and  Abramson \cite{Kuperman} and He and Stone \cite{HeStone}. 
These particular models belong to the   SIS (Susceptible-Infective-Susceptible) class, that is they apply  to diseases which do not confer lifelong immunity. Therefore they are not suited  to describe  typical  childhood diseases. A more problematic feature common to all  these studies  is that  they predict  oscillations with  periods on the  scale of infection and/or immune periods. In contrast, recurrent epidemics of childhood diseases  have periods from a  dozen up to  a few  hundred times the  infection cycle. Notwithstanding, these models are certainly  of great interest as toy models of generic SIS  dynamics.

Whether  recurrent epidemics are governed by an external drive or by the intrinsic nonlinear dynamics has been the focus of intensive debate practically since the dawn of theoretical epidemiology. Here we make a further  contribution to this still unfinished debate  by presenting  a stochastic model that  discards  external factors and considers instead  the heterogeneity  of the contact network. As  reported in a previous publication \cite{VerdascaJTB} the  stochastic implementation of  SIR dynamics in a small-world network can  describe  the onset of  epidemic cycles in a  population, without considering any exogenous factors such as seasonal forcing or immigration of infectives from outside. In our model, infection takes place through a combination of local rules  and long-range contacts, generating  a dynamic  small-world network. In sharp contrast to the  previous studies in the same vein \cite{Johan1,Johan2,Kuperman}, we observe the emergence of  a characteristic time scale which is not that of birth (replenishment of susceptibles) neither  is it related in any trivial way to  the  period of infection.


The results in this paper are arranged in two main parts. The first one is devoted to the SIR implementation. A first set of simulations of long-term behaviour  shows that  the network topology has a strong   impact both  on the amplitude of oscillations and the level of persistence. Then, we consider the evolution of an epidemic in a closed population, with birth and death rates set to zero. We show that the basic reproductive rate, $R_0$, increases from its  minimum value in a regular lattice with local contacts only up to the maximum, mean-field value, as the percentage of long-range infection is increased from zero to one. In the second part of the paper we present simulations of the   more realistic SEIR version.  We calculate the period and amplitude  of oscillations, $R_0$, as well as the average age at infection for realistic demographic and etiological  parameters  corresponding to  measles, rubella and  chickenpox.

The detailed structure of the paper is the following: in Section 2  we  describe the algorithm in detail; in Section 3 we present the  results of the SIR model  and in Section 4 those of the SEIR model. This is followed by a Discussion and, finally, the Conclusion.
 

\section{PCA Model}

\subsection{General description}

Here we describe the probabilistic cellular automata (PCA) implementation of the SIR model that  has been briefly outlined in a preceding paper \cite{VerdascaJTB}.

Individuals live  on a square lattice of $N = L \times L$ sites.  The bonds between sites  are connections along which the infection may spread to other individuals. Infection proceeds either locally, within a  prescribed neighbourhood, or  through a link established at random between any two individuals. We introduce a small-world  parameter $p_{\mbox{\scriptsize SW}}$ defined as  the fraction of attempts at  infection carried out  through a random link; $p_{\mbox{\scriptsize SW}} = 0$ corresponds to a  regular lattice where each individual  contacts with his $k$ nearest  neighbours only,  while $p_{\mbox{\scriptsize SW}}=1$ corresponds to a random network. For intermediate values of $p_{\mbox{\scriptsize SW}}$ the network of contacts is neither fully ordered nor completely random.



\subsection{Algorithm}

We choose first between  {\em birth}, {\em death} and {\em  infection} events, the latter being either {\em local} or {\em long-range}  with respective  probabilities  $p_{\mbox {\scriptsize inf}}^{loc} = (1 - p_{\mbox{\scriptsize SW}})\beta_0$ and $p_{\mbox {\scriptsize inf}}^{lr} = p_{\mbox{\scriptsize SW}}\beta_0$. (Note that we call long-range link any connection established at random to any node on the lattice,  not only those that connect to sites lying  outside the established neighbourhood).  The total population number if fixed, therefore  $p_{birth} \equiv p_{death}$. $\beta_0 = p_{\mbox {\scriptsize inf}}^{loc} + p_{\mbox {\scriptsize inf}}^{lr}$ is the total probability of an attempt at infection while $(1 - p_{birth} - p_{death} - p_{\mbox {\scriptsize inf}}^{loc} - p_{\mbox {\scriptsize inf}}^{lr})$ is the probability that nothing happens.  There is no  restriction associated to the fact that the  sum of  probabilities cannot exceed one  because  
it is  possible to attribute any weight to any one  of the events - birth, death or infection -  by a suitable choice of the time scale; the probability of  not realizing any event   will change accordingly and  a sweep through the lattice will simply correspond to a  different time unit. In one time unit, or PCA step, we  perform $N$ attempts to realize one given event, as follows: we generate a  random number $r$ uniformly  distributed between $0$ and $1$. If $0 < r \le p_{birth}$ we make an attempt to realize a birth event; if   $p_{birth}  < r  \le p_{birth} + p_{death}$ an individual picked at random will die; if $p_{birth} + p_{death} < r  \le p_{birth} + p_{death} +  p_{\mbox {\scriptsize inf}}^{loc}$ an attempt at local infection is made while if  $p_{birth} + p_{death} +  p_{\mbox {\scriptsize inf}}^{loc} <   r  \le p_{birth} + p_{death} +  p_{\mbox {\scriptsize inf}}^{loc} + p_{\mbox {\scriptsize inf}}^{lr}$ the attempt will be  long-ranged. Finally, if $p_{birth} + p_{death} +  p_{\mbox {\scriptsize inf}}^{loc} + p_{\mbox {\scriptsize inf}}^{lr}  <   r \le 1$, we just carry on and generate a new random number. The realization of the events is the following:

\begin{enumerate}
\item Attempt at local infection or long-range infection: First  choose a site  $i$ at random; Then,  

\begin{enumerate}
\item If that site is occupied by a recovered ({\it R}) individual or an infectious ({\it I}) individual, do nothing.

\item If it is occupied by a susceptible ({\it S}) then choose another site  $j$ from   a list of  $k$ possible neighbours, in the case of local infection, or from all of the $N$ sites, with equal probability, in the case of long-range infection. In the simulations presented here, the local range comprises  nearest neighbours, next nearest neighbour and third nearest neighbours ($k=12$). If,

\begin{enumerate}
\item Site $j$ is occupied either by another susceptible or a recovered individual: do nothing.

\item The site  is occupied by an infective: The first individual becomes infected. 
\end{enumerate}

\end{enumerate}

\item A death event is chosen. Then one picks an individual at random who  dies irrespectively of his present state. The probabilities of death are then proportional to the density of  {\it S}, {\it I} and {\it R} individuals. Susceptible and infective individuals who die become recovered; recovered individuals remain in that same class.

\item Birth event:  one looks at the lattice (at random) for a recovered individual. Once found, that individual becomes susceptible. The trial  only ends when one actually finds a recovered individual, so that the birth rate, as it should, is independent of any density. 
\end{enumerate}

\noindent  After a  time during which he stays  infectious to others, the individual recovers. He becomes immune for life and cannot be infected again.  For childhood diseases these periods vary only by a small amount among individuals within a population.  So, it is more realistic to assume a constant infectious period - deterministic recovery - than a  constant probability of recovery leading to an exponential  distribution of infectious periods \cite{Lloyd01}. We model deterministic recovery  by associating  a counter $n_r$ to every individual.  
Upon infection,  the counter for that individual is updated at each time step, $n_r \rightarrow n_r + 1$. At each PCA step  we  move to the recovered class  those  infectives for which the counter has  reached the fixed infectious period  $\tau$, and reset their counter. Stochastic recovery, on the other hand, is modelled by a Poisson process: we add a new event to the list above -- a  recovery event -- taking place with probability $\gamma$. When such event is chosen,  one picks an individual at random; if infective, that individual is moved into the recovered class, otherwise nothing happens.

The SEIR model comprises one additional class, the exposed ({\it E}). After being infected, the individuals enter a latency period during which they cannot be re-infected yet  they are unable to  transmit the infection to others. In our simulations the individuals  move deterministically from the exposed to the infectious class after $\tau_{lat}$ steps and then stay infectious for $\tau_{inf}$.

One important difference between the model and the population dynamics  in developed countries is  the implicit assumption  of the so-called Type II mortality, where individuals die with equal probability  independently of their age, as opposed to Type I mortality where all individuals live up to the same age and then die \cite{AM}.

\section{Results:  SIR model}

\subsection{Persistence}

\begin{figure}[ht]
\centerline{\epsfxsize=2.8in\epsfbox{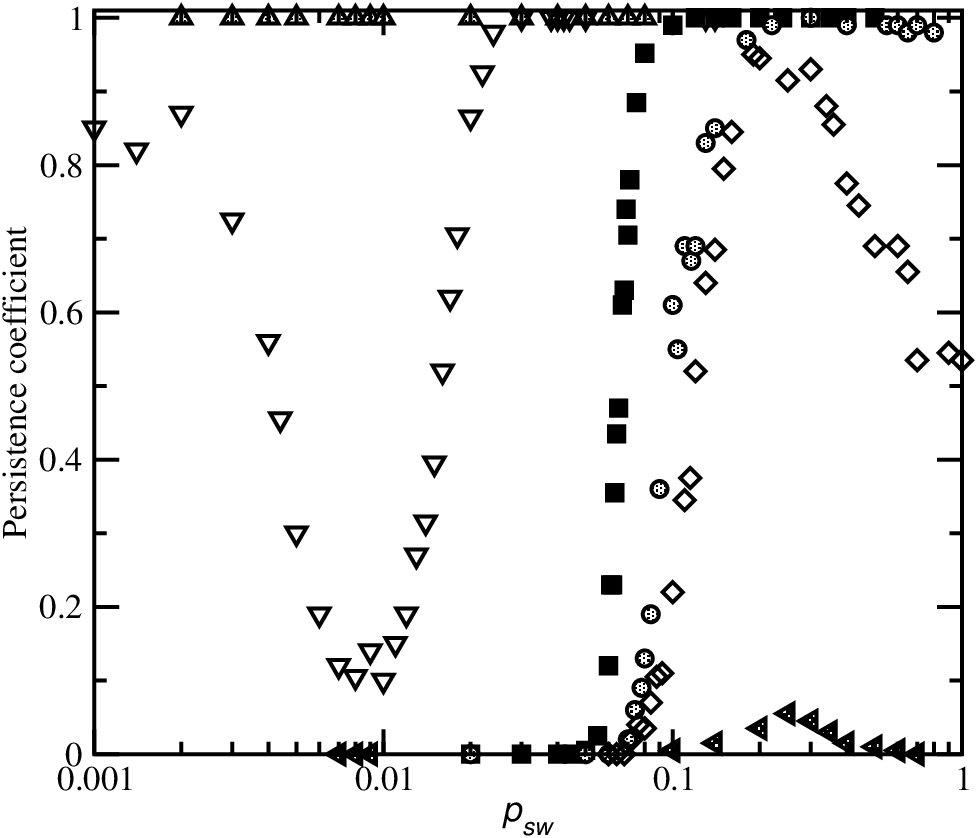}}
\caption{\small{Persistence coefficient, defined as the fraction of  runs that attain a prescribed time $t_{max} = 20 000$ of a total of $n = 100$ runs started, calculated on a   $400 \times 400$ lattice with deterministic recovery. The initial fraction of susceptibles was $s_0 = 0.165$. The probability of infection is $\beta_0 = 0.66$ and the infectious period  $\tau = 16$. The different curves correspond to the  following  birth rates: $\mu = 0.0002$ (triangles left), $\mu = 0.0004$ (diamonds), $\mu = 0.0006$ (circles), $\mu = 0.001$ (squares),  $\mu = 0.002$ (triangles down) and  $\mu = 0.005$ (triangles up). }} \label{persist2}
\end{figure}

Recurrent epidemics can persist in finite populations because of the finite birth rate that allows for the renewal of susceptibles. The way in which persistence varies with the small-world parameter $p_{\mbox{\scriptsize SW}}$ depends crucially  on the rate at which fresh susceptibles are supplied by birth. As shown in Fig. \ref{persist2},  for intermediate values of $\mu$ the persistence coefficient is zero at low $p_{\mbox{\scriptsize SW}}$, and approaches one over a narrow range of $p_{\mbox{\scriptsize SW}}$. 
At values $\mu \lessapprox 0.001$ a maximum starts to develop. Barely noticeable at $\mu = 0.0006$, it is already quite pronounced at $\mu =  0.0004$.  At  $p_{\mbox{\scriptsize SW}}= 1.0$, only about half the runs survive up to  the maximum ascribed time, whereas within the range  $p_{\mbox{\scriptsize SW}}= 0.2 - 0.3$ more than 90 \% of the runs reach $t_{\mbox{\scriptsize max}}$. The relevant fact is that at   low enough birth rates there is an optimal value of $p_{\mbox{\scriptsize SW}}$ for the disease to persist in finite populations.  Conversely, for $\mu \gtrsim 0.001$ the persistence coefficient shows a minimum at finite  $p_{\mbox{\scriptsize SW}}$. The logarithmic scale in  Fig. \ref{persist2} enables us to highlight this remarkable symmetry of behaviour  but here we must note that  what happens at   high birth rates is irrelevant for childhood diseases. Indeed, for $\tau$ given in days (which is not unreasonable) we get   a birth rate of $\mu = 0.002$ day$^{-1}$ corresponding to an average lifespan of 1.4 years! Even for the lowest $\mu$ with which  we can still  observe a reasonable  level of persistence, the lifespan would only be around  $7$ years. This feature is a distinguishing property of the PCA implementation of the SIR model, namely that to obtain realistic patterns of persistence one has to choose birth rates at least one order of magnitude higher than the real values. We shall see in Section 4, that the adoption of the more realistic  SEIR dynamics permits  to avoid this problem.

\begin{figure} [ht]
\centerline{\epsfxsize=2.4in\epsfbox{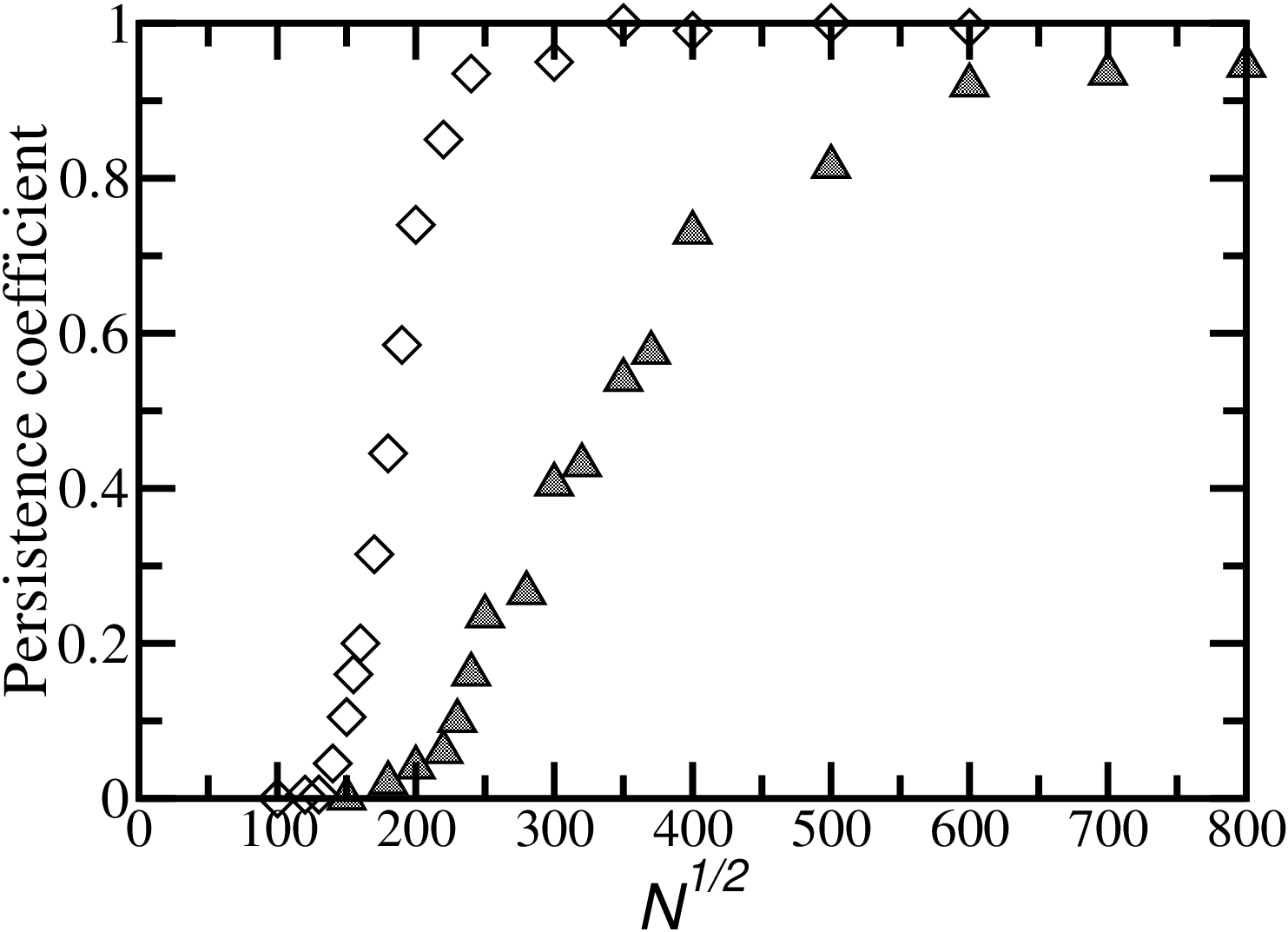}}
\caption{\small{Persistence coefficient as a function of lattice size for $p_{\mbox{\scriptsize SW}}=0.07$ (triangles) and $p_{\mbox{\scriptsize SW}}=0.1$  (diamonds) at $\mu =  0.001$. The other parameters are the same as in Fig. \ref{persist2} except $n = 200$.}} \label{persistN}
\end{figure}

The fact that persistence depends  on the fraction of long-range contacts implies that the Critical Community Size (CCS)  also depends on the structure of the contact network. In order to demonstrate how the PCA can be used to estimate the Critical Community Size, and how the CCS depends on the fraction of long-range infection we  choose two examples, at $p_{\mbox{\scriptsize SW}}  = 0.07$ and $p_{\mbox{\scriptsize SW}}  = 0.1$, with a birth rate  $\mu = 0.001$ (squares in  Fig. \ref{persist2}).
While  at  $p_{\mbox{\scriptsize SW}}=0.1$ there is  a sharp rise in persistence,  at  $p_{\mbox{\scriptsize SW}} = 0.07$  a much smoother curve is obtained   (Fig. \ref{persistN}).  The population size at which the persistence rises above 50 \% can be taken  as a  estimate  of the CCS 
but  we might as well choose a different threshold. 
\vspace{2mm}

\subsection{Amplitude of oscillations}

\begin{figure} [ht]
\centerline{\epsfxsize=3.6in\epsfbox{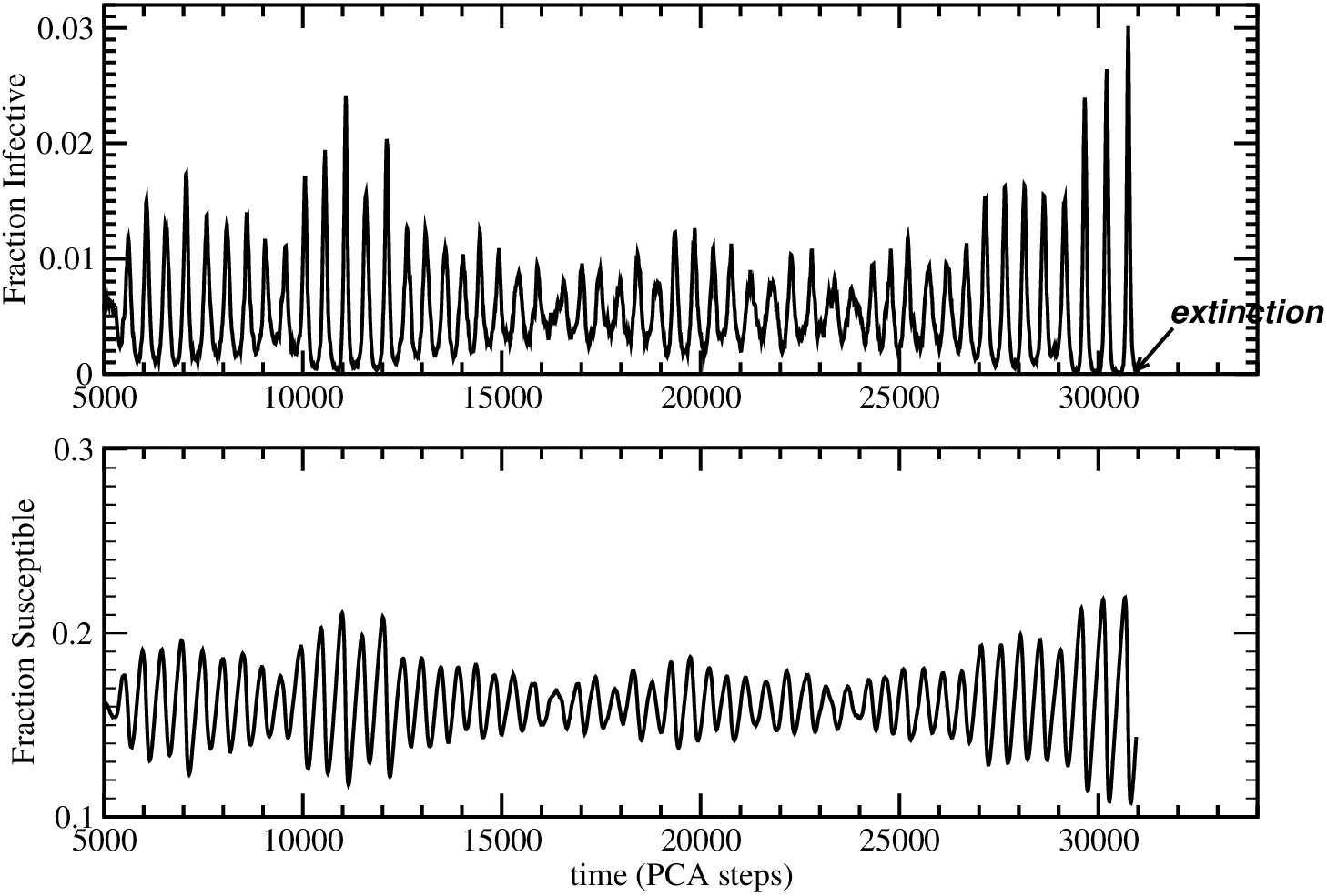}}
\caption{\small{Typical evolution of the fraction of infectives (top)  -- number of infective individuals divided by the system size $N$ --   and the fraction of susceptibles (bottom), obtained with  the SIR PCA model implemented on a  small-world network of  $N = 160000 \equiv 400 \times 400$ individuals. In the simulations,    90 \% of the contacts are  local and the remaining 10 \% long-range ($p_{\mbox{\scriptsize SW}}=0.1$).   Recovery was stochastic  with a probability $\gamma = 0.0625 \equiv 1/16$. The other parameters are $\mu = 0.0004$ and  $\beta_0 = 0.66$. The system oscillates for  more than 50 cycles before extinction occurs after about 30950 PCA steps.}} \label{longsir}
\end{figure}

Fig. \ref{longsir} shows a typical output of the  SIR model implemented on  a fairly large lattice. The time series shown are for the susceptible and infective fraction.   It is impossible to run  the  three-state SIR implementation  for a long time using  realistic demographic and etiological parameters. Extinction is almost certain to occur after only a few cycles. Thus the values of the birth rates used here are much larger that real ones, again  for  a $\tau$   of a few days.

\begin{figure} [ht]
\centerline{\epsfxsize=3.6in\epsfbox{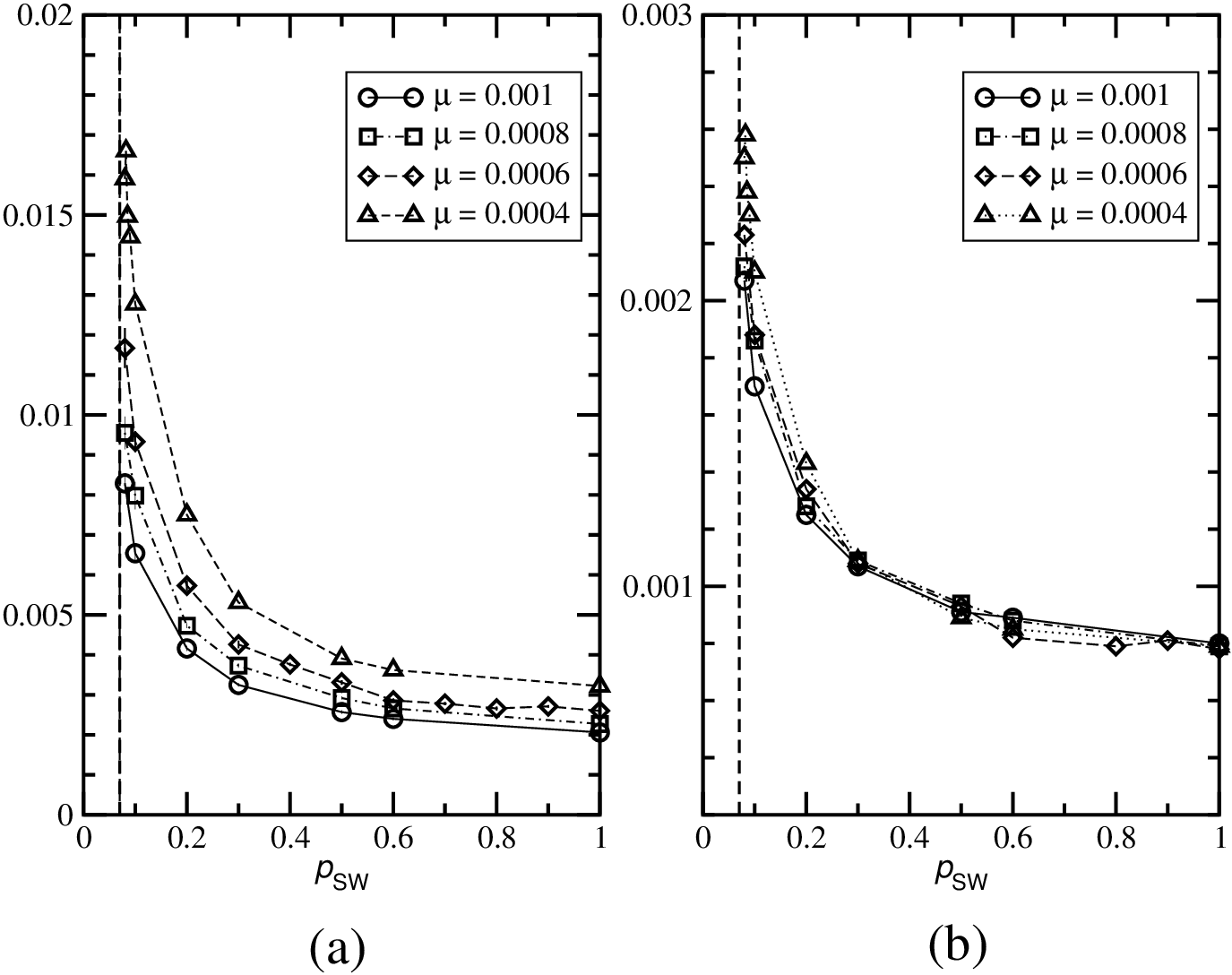}}
\caption{\small{RMS amplitude of oscillations in the fraction of susceptibles (a) and infectives (b) as a function of the fraction of long-range contacts for decreasing values of the birth rate. Other parameters the same as  in Fig. \ref{longsir}. Stochastic recovery. Each point represents an  average over 10 runs started from different initial conditions.   }} \label{rmsamps}
\end{figure}

When  $p_{\mbox{\scriptsize SW}}$  is small, the oscillations have  large amplitudes  and are  strongly  synchronized.  Keeping all other parameters unchanged but  setting   $p_{\mbox{\scriptsize SW}}=1.0$ we observe that  oscillations become  much more irregular and their amplitude is   considerably diminished. 
In Fig. \ref{rmsamps} we plot the root mean square (RMS) amplitude of the oscillations in the fraction of susceptibles (a) and infectives (b) as a function of the birth rate for different values of  $p_{\mbox{\scriptsize SW}}$. As $\mu$ is decreased the RMS amplitudes are amplified and this trend   is the more pronounced the smaller the value of   $p_{\mbox{\scriptsize SW}}$.

Looking at the data from a different perspective now, one detects quite clearly the enhancement of stochastic fluctuations growing  into fully developed oscillations as the relative weight  of  long-range infection is decreased. The smaller the birth rate the more accentuated the effect becomes: at $\mu=0.001$  the amplitude of the susceptible oscillations at $p_{\mbox{\scriptsize SW}}=0.08$ is approximatively four  times that of the homogeneous mixed population while at $\mu=0.0004$ the  same ratio is  almost five. The increase in the amplitude of oscillations in the fraction of infectives (Fig. \ref{rmsamps} (b))  follows that of susceptibles but it is less marked.

\subsection{Effective transmission rate}

Diseases do not spread very effectively on lattices when only local contacts are allowed because infective individuals tend to interact  mostly with other already infected individuals. To evaluate the impact of the aggregation of infectives and susceptibles into clusters  on the spread of the disease brought about by the local contact rules we can  estimate the transmission rate directly from the simulations. This is  done by observing that once the transients have died out,  the mean  number of new infections that take place in a short  time interval $\Delta t$, $nni_{\Delta t}$,   is approximatively proportional to the product of the mean number of infective individuals by the mean number of susceptibles during that same time interval: 
\begin{equation}
nni_{\Delta t}  \sim \overline{I}_{\Delta t} \times  \overline{S}_{\Delta t}. \label{nni}
\end{equation}
The sensitive issue here  is  to choose a suitable value for  $\Delta t$, which must be  much smaller than the average period of oscillations or otherwise  we would not get an instantaneous measure of transmission, but at the same time considerably larger than the infection period so that stochastic effects  get averaged out. There is no recipe to pick up the right value, but  choosing a  $\Delta t$ of a  few dozen time steps usually guarantees that  the proportionality  (\ref{nni}) is verified. In that case,  we can define an  {\em effective} transmission rate as:
\begin{equation}
\beta_{\mbox {\scriptsize eff}} (t) = {{nni}_{\Delta t}(t)  \over   \overline{I}_{\Delta t}(t)  \overline{S}_{\Delta t}(t) }. \label{effbeta} 
\end{equation}

\noindent
 This instantaneous  transmission rate   fluctuates wildly on the time scale of the birth rate but once the transients have died out we can calculate the temporal mean. This (averaged) effective transmission rate stays  always below the mean-field transmissibility  $\beta_0$. Indeed, by aggregating infectives and susceptibles into clusters, the structure of local infection  on a  {\em regular} network structure acts to  keep the number of  contacts that can  actually  result in transmission, namely those  between a susceptible and an infective, well below the level that    would result  if the individuals  were homogeneously distributed on the lattice irrespectively of their disease status. Although, as we shall see below, the dynamic small-world structure of contacts exhibits some features of the well-mixed situation, {\em locally} the structure of contacts remains highly clustered. 
As long as  infected individuals remain in contact mostly with other already infected individuals the progression of the disease stalls but once  an individual belonging to an  infectious cluster establishes a shortcut that  propagates the disease into a region where susceptible individuals predominate, the disease will perhaps get a new boost that will carry it through one more cycle.

\vspace{4mm}

\begin{figure} [ht]
\centerline{\epsfxsize=3.2in\epsfbox{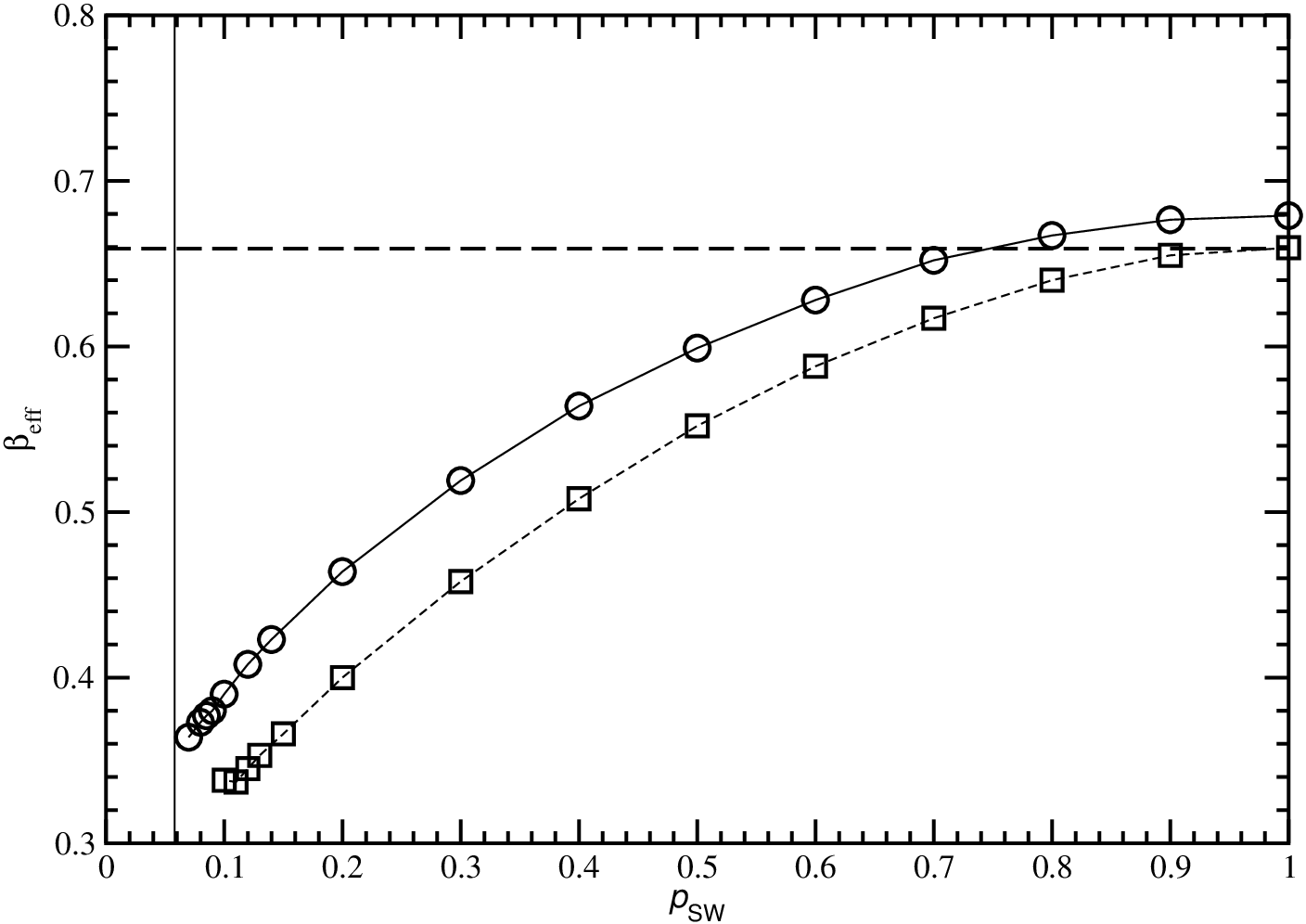}}
\caption{\small{Effective transmission rate  $\beta_{\mbox{\scriptsize eff}}$ vs. the small-world parameter $p_{\mbox{\scriptsize SW}}$. The reduction of  $\beta_{\mbox{\scriptsize eff}}$ that arises both for deterministic (circles) and stochastic (squares) recovery is due to the clustering of infectives and susceptibles. The dashed line indicates the value of the transmissibility, or total probability of realizing an  infection event used in the simulations: $\beta_0 = 0.66$. Note the almost perfect agreement with the mean-field result  $\beta_{\mbox {\scriptsize eff}} \equiv \beta_0$ observed at $p_{\mbox{\scriptsize SW}} = 1.0$ for stochastic recovery. The birth rate is $\mu = 0.0006$,  the infectious period for deterministic recovery $\tau = 16$ and $\gamma = 0.0625 \equiv  1/\tau$ for stochastic recovery. Each point represents an  average over 10 runs. }} \label{betavsp}
\end{figure}

As shown in Fig. \ref{betavsp}, $\beta_{\mbox {\scriptsize eff}}$ increases smoothly as $p_{\mbox{\scriptsize SW}}$ increases. 
For values of  $p_{\mbox{\scriptsize SW}}$ to the left of  the lower end of  the curves    only a very small fraction of the runs last for  more than a few cycles. With such a small persistence it  becomes impossible  to follow the long term dynamical regime and compute repeated  time averages of non-transient oscillations.  The two curves shown in Fig. \ref{betavsp} correspond to stochastic and deterministic recovery. Their shape is identical, but the value of $\beta_{\mbox {\scriptsize eff}}$  for stochastic recovery always stays below that obtained with  deterministic recovery. The values of  $\beta_{\mbox {\scriptsize eff}}$ for  stochastic recovery converge to the mean-field value for $p_{\mbox{\scriptsize SW}} \rightarrow 1$ with a small discrepancy due  to finite-size effects, whereas for deterministic recovery $\beta_{\mbox {\scriptsize eff}}$  is still about  3 \% above $\beta_0$ in the same limit.  Moreover we  found that the amplitude of oscillations is also larger for deterministic recovery than for stochastic recovery, the differences being small at large $p_{\mbox{\scriptsize SW}}$  but increasing significantly as  $p_{\mbox{\scriptsize SW}}$ is lowered.  This is   in agreement with a previous study of a homogeneously mixed stochastic model,  showing    that  sharp distributions of the infection period result in larger fluctuations than those  with a  smooth distribution \cite{Lloyd01}. In our model this effect actually intensifies when contact structure is introduced.

\subsection{Basic reproductive rate}

It is incontrovertibly  accepted that in order to figure out if a given pathogen will be able  to succeed in a host population the crucial quantity to compute is the basic reproductive rate $R_0$. This is the number of secondary infections produced by an infected individual in a population entirely composed of susceptibles.

The  analysis of steady state solution of  the SIR deterministic ODE's  considering only  ¨weak¨ homogeneous mixing -- meaning that the rate of new infectives is proportional to the total number of susceptibles -- and  Type II survival, gives \cite{AM}:
 \begin{equation}
R_0 = {1 \over \mu A}, \label{R01}
\end{equation}
where $A$ is the average age at infection, in the case of constant population size. Under the more stringent condition imposed by the  mean-field approximation,  one has also \cite{AM}
\begin{equation}
R_0 = {\beta N \over \gamma + \mu}. \label{R02}
\end{equation} 
Note that relations (\ref{R01}) and (\ref{R02}) suppose stochastic recovery.

The  definition of $R_0$ rests on the existence of  a pristine susceptible population, {\em i.e.} it is valid only at vanishing infective fraction. However, it is a consequence of mean-field models and a fact  confirmed  by the analysis of many real epidemics,  that following the introduction of a pathogen in a population consisting  entirely of susceptible individuals  the number of infective individuals grows  exponentially in the early stages:
\begin{equation}
I(t) \sim e^{\Lambda t}, \label{expgrow}
\end{equation}
where $\Lambda =  (R_0 -1)/\tau$. So, with the exception of  incipient epidemics  ($R_0 \rightarrow 1$),  very soon after an epidemic has started from a single infective, the number of infectives is already too large for the definition of $R_0$ to apply. What  can be  measured directly from the PCA in long-time simulations, and in most field studies is rather  $R_t$, the effective reproductive rate at a given time $t$ during the evolution of the disease when  there are also infective and recovered individuals present. Upon the further assumption of ¨weak¨ homogeneous mixing, we can write\begin{equation}
R_{t \rightarrow \infty} = R_0 s^*  \label{rtvss}
\end{equation}
 where $s^*$ is the fraction of susceptibles in the endemic steady-state. If the rate at which susceptible individuals are infected is exactly  balanced by the rate at which new susceptibles are born  then   $\bar{R}_{t \rightarrow \infty} = 1$, where the bar denotes the temporal mean. That is, each primary  infection will produce on average exactly one secondary infection. Under these precise conditions, the basic reproductive rate is simply $R_0 = 1/s^*$.

\begin{figure} [ht]
\centerline{\epsfxsize=3.1in\epsfbox{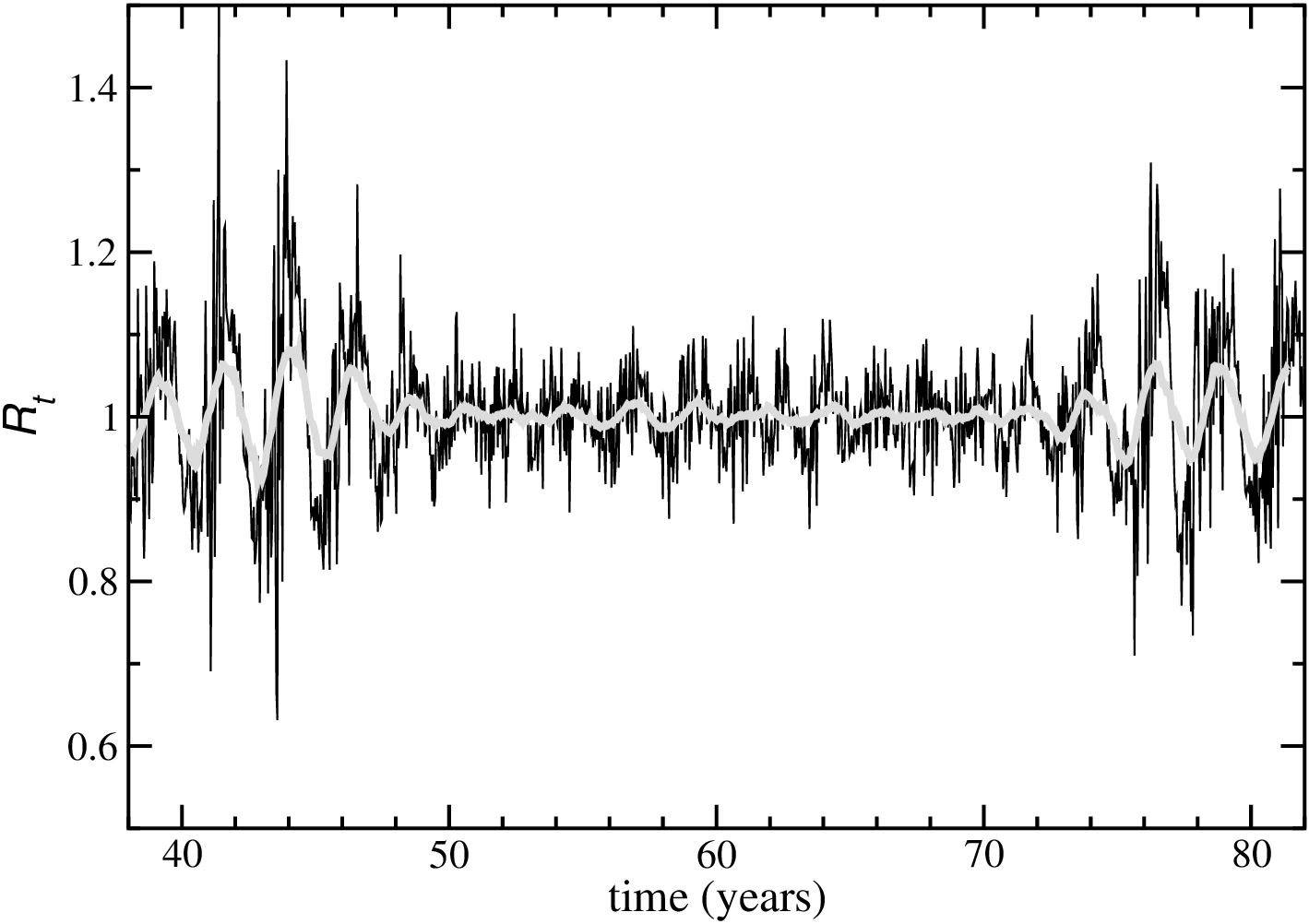}}
\caption{\small{Evolution of the effective reproductive rate in a SEIR simulation. $R_t$ is calculated by counting the number of secondary infections caused by every infective individuals and taking averages. The result is a very noisy time series which nevertheless shows an  underlying structure composed of cycles with the same period as the oscillations in incidence.   Smoothing  the data  by averaging it out with a moving window  brings forward the pattern of oscillations (thick plain curve). In the  SIR model we observe the same effect only the data contains  a lot more noise. }} \label{reprorate}
\end{figure}

The depletion of susceptibles during the epidemic implies that $R_t < R_0$ always. If $R_t$ is maintained below 1 for  sufficiently long, then  the pathogen will become extinct.  However, as we can see in Fig. \ref{reprorate}, $R_t$ calculated directly from the PCA oscillates  around an average value  which, within statistical errors, is exactly one. $R_t$ drops below the self-sustained threshold for half the period of oscillations only to rise again above it in the next half-cycle. 
This result has drastic  implications on the assessment of the risk of  recurrent outbreaks of infectious diseases based on calculations of $R_t$. It is usually assumed that as soon as $R_t$ drops below 1, an epidemic is  on its way to be  contained. Very recently  this criterium was used to judge the outcome of the SARS epidemic \cite{SARSRiley,SARSLip} based on data spanning only a few weeks. In this vein, the  results in  Fig. \ref{reprorate}  should act as a warning particularly when, as it is often happens, one has to  deal with short time series or otherwise incomplete data.

Now  we must make the  distinction between the long-time measurements, using  time averages over (extended) time series of recurrent epidemics  and short-time measurements  of $R_0$.  For the long-term measurements we have different possibilities to estimate $R_0$: We can either i) compute $A$ directly from the PCA by the method described in the next Section and then use Eq. \ref{R01} to evaluate $R_0$. ii) Estimate the effective transmissibility $\beta_{\mbox{\scriptsize eff}}$ as described in Section 3.3 and then compute $R_0$ from  (\ref{R02}) with $\beta \equiv \beta_{\mbox{\scriptsize eff}}$ or iii) use the fact that $R_t = 1$ once an asymptotic regime is attained and estimate  $R_0 = 1/s^*$. 

\begin{figure} [ht]
\centerline{\epsfxsize=3.4in\epsfbox{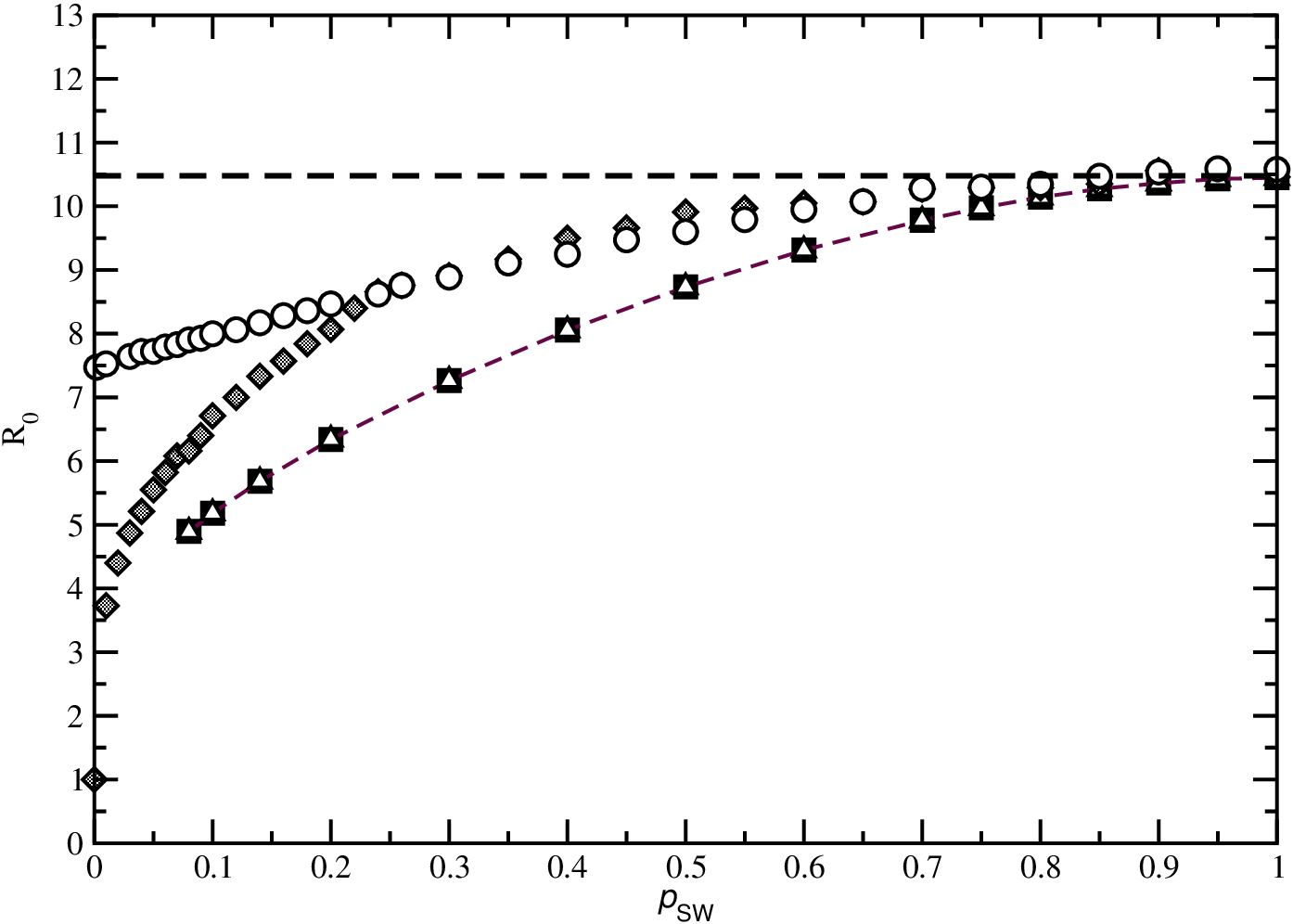}}
\caption{\small{$R_0$ as a function of the small-world parameter $p_{\mbox{\scriptsize SW}}$. The two upper curves were calculated from simulations in a closed population ($\mu = 0$),  from the final size equation (\ref{FSE}) (circles) and by   fitting  exponential growth using Eq. (\ref{expgrow}) (diamonds). At  $p_{\mbox{\scriptsize SW}}$ the number of infectives grows in time as a power law and therefore $\Lambda \equiv 0$ implying $R_0 \equiv 1$. The curves obtained from the analysis of  long-term dynamics  using $R_0 = 1/s^*$ (squares) and from $A$ and  Eq. (\ref{R01}) (triangles) are almost indistinguishable.  In the limit of a random network the four curves converge to $\beta_0 / (\gamma + \mu)$, indicated by the dashed line, with a small discrepancy attributed to final size effects. SIR model in a $400 \times 400$ lattice with stochastic recovery; parameters are the same as in Fig. \ref{betavsp}.}} \label{r0vsp}
\end{figure}

As an alternative to perform  lengthy simulations of recurrent epidemics for the purpose of computing $R_0$, we can  run the PCA  in a closed population. Discarding births and deaths by setting $\mu = 0$,  we can  iv) compute the $R_0$ from the final size equation (FSE) \cite{Diekmann}
\begin{equation}
\ln (s_\infty) = R_0 (s_\infty - 1) + \ln(s_0), \label{FSE}
\end{equation}
where $s_\infty = S_\infty/N$, with  $S_\infty$  the number of susceptibles left once the epidemic has died out. The initial fraction of susceptibles is $s_0 = (N - 1)/N$ since every new run  begins with a single infective. The last form of computing $R_0$ that we considered was to v) fit the  exponential growth law, Eq. (\ref{expgrow}), to the early stages of evolution of an epidemic ravaging a wholly susceptible population.
The results are presented in Fig. \ref{r0vsp}. They show how  $R_0$ can be used  to assess  the departure from mean-field behaviour as the fraction of long-range infection is decreased, and also that  in a structured population with the characteristics of a dynamic small-world, the magnitude of  $R_0$ is always below the value that would be  obtained  in an   homogeneously mixed population. 
Different procedures of estimating $R_0$ that would give the same result under the mean-filed approximation will now yield markedly distinct values. This is because they still appeal to an approximation  at some stage but not all at the same stage. Undoubtedly, this is a  blow to the worth of $R_0$  evaluated in practical situations. Indeed, with the exception of  contact tracing in the very early stages of an epidemic, which is a  very difficult task to perform,  $R_0$ can only be determined through indirect methods. Our simulations show that  evaluating $R_0$ from different sources, for instance from  $A$ obtained from serological studies, or from  the equilibrium number of susceptibles obtained  from historic time series can lead to {\em  intrinsically} different results as a consequence of  the heterogeneity of contacts. 

\vspace{2mm}

\subsection{Average Age at Infection}

The moment of their lives  when susceptible individuals acquire infection is a very important epidemiological quantity. The average age at infection, $A$  is inferred from serological surveys and  can be compared to the output of epidemiological models.

We  evaluate the impact of network structure on  $A$  by calculating it   directly from the simulations as follows:  a counter is associated to every susceptible and the moment when this individual is infected is recorded. Averaging over every single individual who has become infected so far   one obtains a  quantity that  shows a slow but unremitting trend towards asymptotic behaviour characterized by small, rapid  fluctuations around a steady-state value. Once the long term evolution appears to  stabilize we compute the time average.

\begin{figure} [ht]
\centerline{\epsfxsize=2.6in\epsfbox{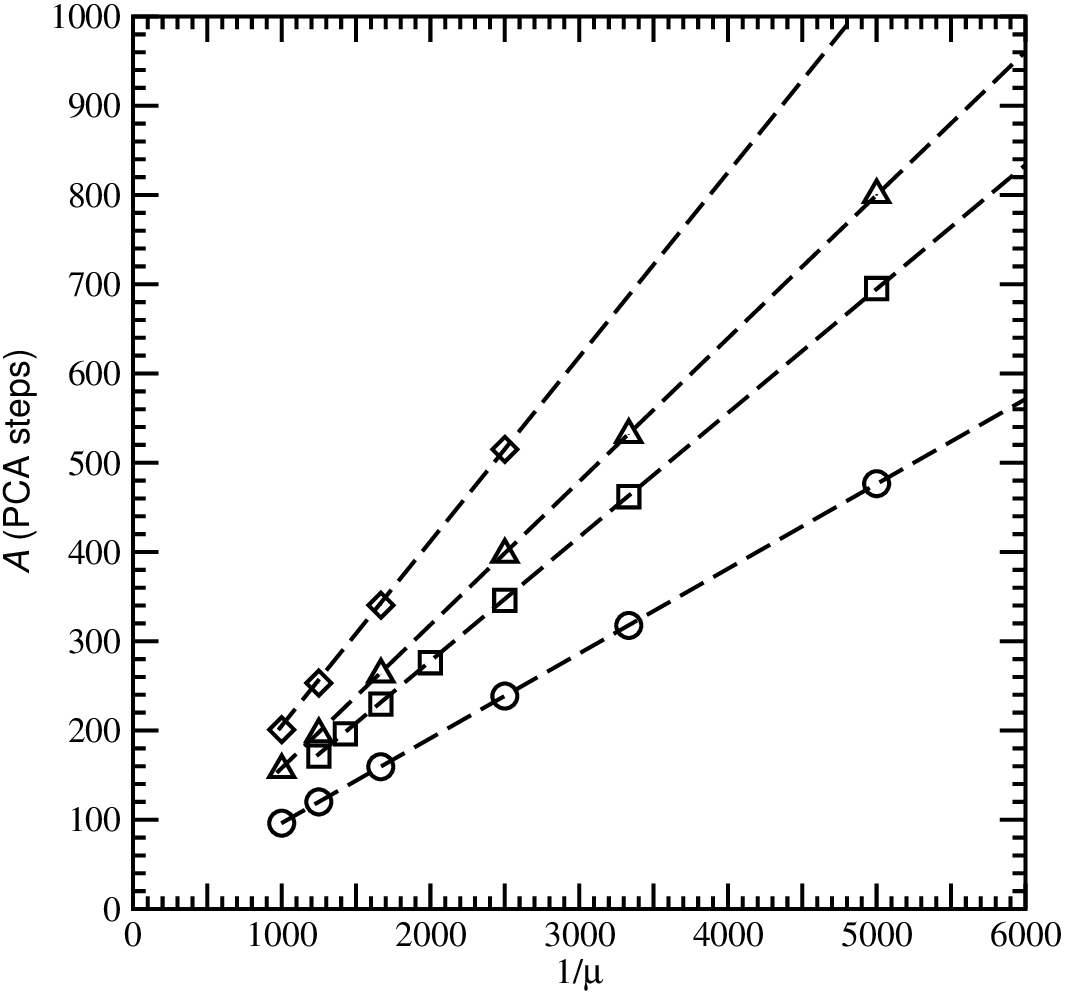}}
\caption{\small{The average age at infection measured in PCA steps as a function of life expectancy $1/\mu$ in the same units, for $p_{\mbox{\scriptsize SW}}=1.0$ (circles), 0.3 (squares), 0.2 (triangles) and 0.08 (diamonds). The slope of the lines gives $1/R_0$. The parameters are $\beta_0$ = 0.66, $\gamma = 0.0625$; the initial infective fraction was $s_0 = 0.165$ and 50 infectives were present on the lattice. The values of $A$ were obtained from a temporal average of data sampled from $t = 10000$ to $t=50000$ at 10 steps interval  and further averaging over 10 realizations starting from different initial conditions.   }} \label{aaivsmu}
\end{figure}

The average age at infection was found to be linear in $1/\mu$ down to only \break 8 \% of long-range contacts. Below that value of $p_{\mbox{\scriptsize SW}}$ persistence was too low to obtain meaningful averages. The  slope of the lines in Fig. \ref{aaivsmu} gives $1/R_0$. 
Moreover, the ratio $r  = R_t A \mu / s^*$ varies  from   $0.9999 \pm  0.0002$ at $p_{\mbox{\scriptsize SW}}=1.0$ to   $r = 0.9994 \pm 0.0001$ at  $p_{\mbox{\scriptsize SW}}=0.08$  showing that mean-field relations hold to an excellent approximation down to surprisingly low values of  $p_{\mbox{\scriptsize SW}}$. Deviations to mean-field behaviour do occur but are only slight even in relatively small populations.  From the data in   Fig. \ref{aaivsmu}  we can  extrapolate the average age at infection at low birth rates. Setting the time scale by taking  $L \equiv  1/\mu = 61$ years we obtain $A = 6$ years for $p_{\mbox{\scriptsize SW}}=1.0$, $A = 7.7$ years for  $p_{\mbox{\scriptsize SW}}=0.3$  and $A = 10.2$ years for  $p_{\mbox{\scriptsize SW}} = 0.08$ showing that the average age at infection increases significantly when we consider a local, clustered network of contacts.

\section{Results: SEIR model}

\subsection{Epidemic cycles}

\begin{figure} [ht]
\centerline{\epsfxsize=3.6in\epsfbox{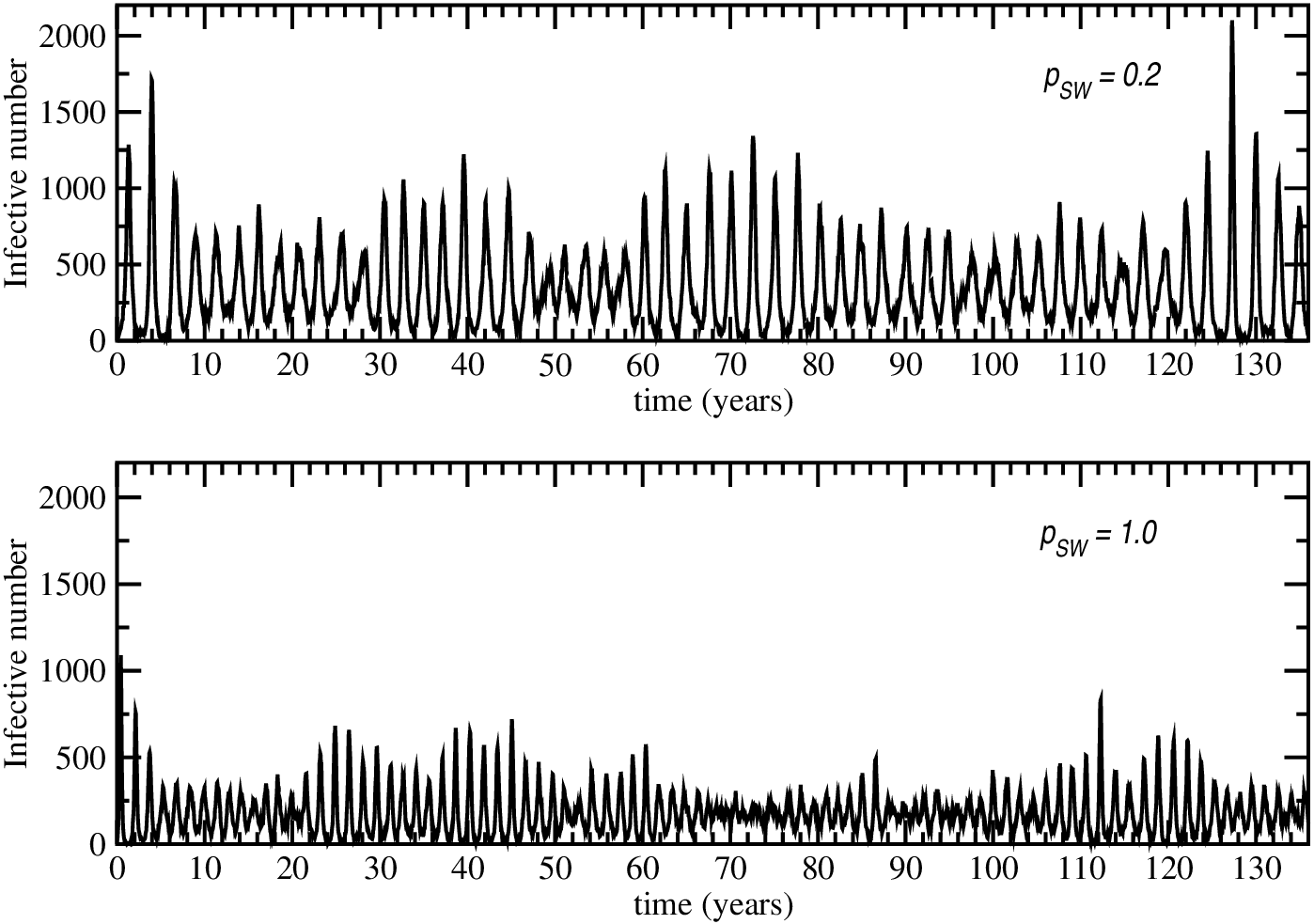}}
\caption{\small{ Time series of the number of infectives obtained from simulations of SEIR dynamics in a $1000 \times 1000$ lattice with $\mu = 1/61$ yr$^{-1}$, $\tau_{lat} = 6$ days, $\tau_{inf} = 8$ days and $\beta_0 = 3.92$ day$^{-1}$.}} \label{measles1}
\end{figure}

We now present numerical  simulations of the  more sophisticated  SEIR model, with etiological and demographic parameters corresponding   to  measles, rubella and  chickenpox, in developed countries, in the pre-vaccination era.  
Just as in the SIR case we observe sustained oscillations in incidence. The oscillations obtained from  the SEIR model  are less coherent than those obtained with the SIR version. 
Their aspect is actually closer to the observed time series.

But the most consequential finding of the SEIR simulations   is the realization that one can only obtain sustained  oscillations with  amplitudes compatible with the existing data records, and for realistic values of the model  parameters -- ({\em e.g.} life expectancy, latency and infectious periods) --  if one chooses values of $p_{\mbox{\scriptsize SW}}$ in the small-world region \cite{Watts}.

To illustrate this feature,  we show in Fig. \ref{measles1} two  time series for measles,  one  obtained for  $p_{\mbox{\scriptsize SW}}= 0.2$ and the other for  $p_{\mbox{\scriptsize SW}} = 1.0$. The amplitude of the oscillations is  almost double in the small-world network.

\begin{figure} [ht]
\centerline{\epsfxsize=3.8in\epsfbox{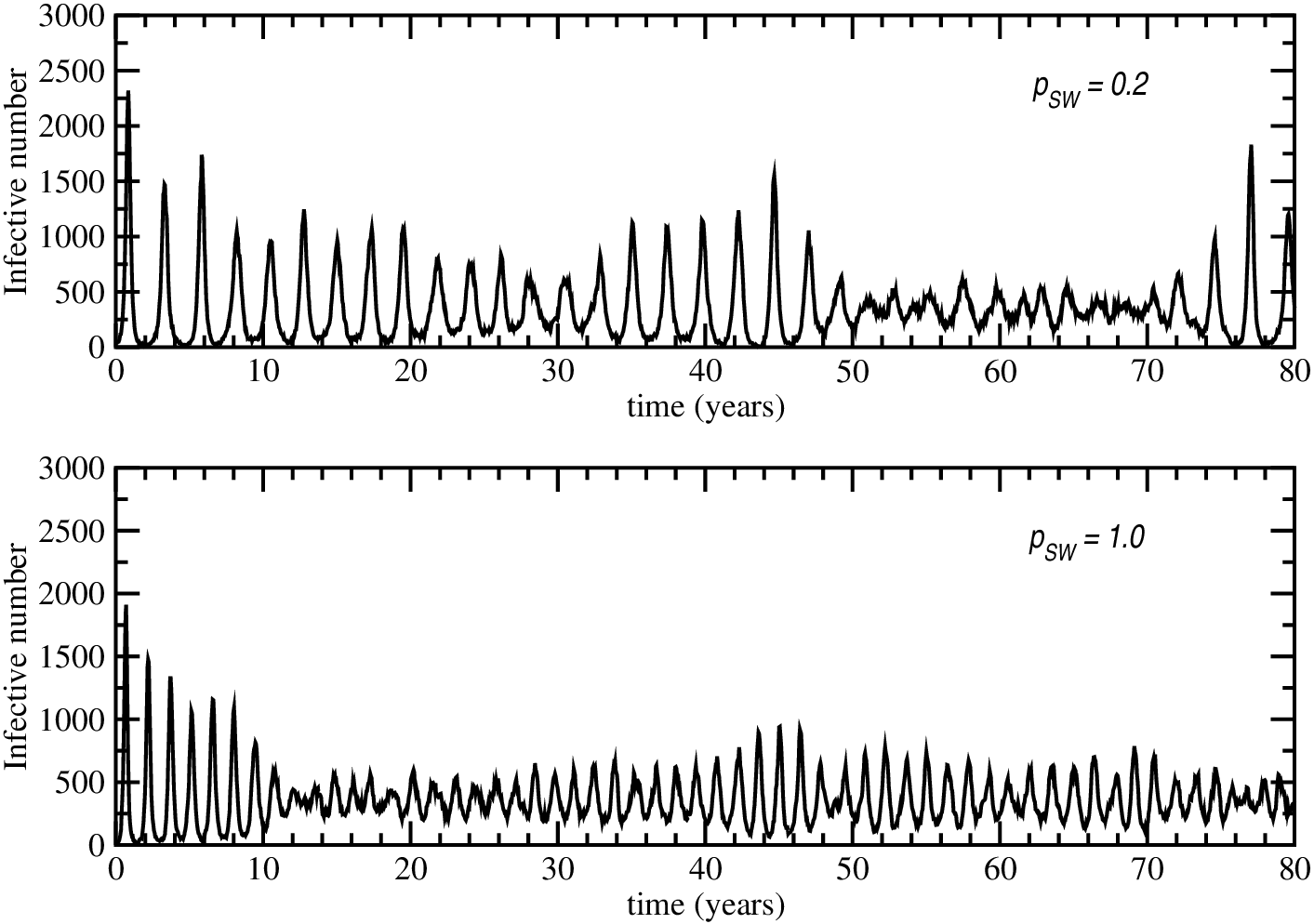}}
\caption{\small{Same parameters as in  Fig. \ref{measles1} except that now $\beta_0 = 4.75$ day$^{-1}$. The period of oscillations shrinks when $\beta_0$ increases. Note the change of scale in both axes with respect to  Fig. \ref{measles1}.}} \label{measles2}
\end{figure}

For  $p_{\mbox{\scriptsize SW}} = 1.0$ the frequency distribution is peaked around   1.5  years while for  $p_{\mbox{\scriptsize SW}} = 0.2$ the peak is at about 2.5 years.
For measles, almost every data record in developed countries  points to cycles of almost exactly  2 years, in between those two values.  Agreement  with the observed periods can be improved,  but only to some extent,  by fine tuning the transmissibility.
Indeed, increasing $\beta_0$  has the effect of decreasing the period 
making it more in line with the observations. The resulting time series are shown in Fig. \ref{measles2}. Nevertheless, we must stress that  $\beta_0$ is nothing like a free parameter. First of all it cannot be changed at will in order to tune the period because the amplitude of oscillations also changes and  the average  size of outbursts  must remain comparable to those observed in cities with the same population size \cite{VerdascaJTB}. Secondly, since the transmissibility of  typical   childhood diseases must be  similar (see Section 4.3 below), differences in period, average age at infection, etc. between them must be achievable with values of $\beta_0$ of  the same order of magnitude and  respecting the expected  infectious rank of those diseases.  Finally, and most importantly,  although $\beta_0$ is  a parameter that can  only be very loosely inferred from the data, there are indirect estimates of  $\beta_{\mbox {\scriptsize eff}}$  from $R_0$ and $A$, using mean-field relations, that set an order of magnitude for $\beta_0$.

\subsection{Estimates of $A$ and $R_0$}

We have  computed the average age at infection obtaining   $A = 1.6$ years for  $p_{\mbox{\scriptsize SW}} = 1.0$ and $3.6$ years for $p_{\mbox{\scriptsize SW}}= 0.2$ for the simulations  shown in  Fig. \ref{measles2}, and $A = 1.9$ years for  $p_{\mbox{\scriptsize SW}} = 1.0$, and $4.1$ years for $p_{\mbox{\scriptsize SW}}= 0.2$ for the simulations in Fig. \ref{measles1}. While the latter is just barely above the lower bound of the interval commonly accepted to correspond to  measles data -- between 4 to 6 years -- the values obtained with the  homogeneous mixed population  lie notably  outside the realistic range. A further refinement consisting in the inclusion of  protection by maternal antibodies in newborns could increase both values by 3 to 9 months  putting the result for $p_{\mbox{\scriptsize SW}}= 0.2$ well inside  the interval of realistic values. However, this is expected to  have a deleterious repercussion on the period of oscillations,  increasing it above what is consistent with data records for measles.



For the simulation at  $p_{\mbox{\scriptsize SW}}= 0.2$ in Fig. \ref{measles1}, the average fraction of susceptibles was $s^* = 0.0657$ giving a value for $R_0 = 1/s^*$ of  $15.2$ while for $p_{\mbox{\scriptsize SW}}= 1.0$ we get  $s^* = 0.0312$ yielding $R_0 = 32.1$. Whereas the former  lies  inside the reported  range of $R_0$ for measles, $14 - 18$ \cite{AM}, the latter is way off range. 
For the SEIR model, the mean-field expression for $R_0$  is slightly more complicated than in the SIR case \cite{AM}:
\begin{equation}
R_0 = {\beta  \over \tau_{inf}^{-1} + \mu } \left ( {\tau_{lat}^{-1}  \over \tau_{lat}^{-1} + \mu } \right ). \label{R0_SEIR}
\end{equation}
Solving  for $\beta$ and  using the values of $R_0$ calculated above  we obtain  $\beta = 4.0$  and $\beta = 1.9$ for $p_{\mbox{\scriptsize SW}}= 1.0$ and $0.2$ respectively. While the former is  close to the $\beta_0$ used in the simulations, the latter  is practically one  half.  The reason for this is that the transmission rate  computed from Eq. (\ref{R0_SEIR}) is rather the   $\beta_{\mbox {\scriptsize eff}}$   introduced in Section 3.3, and only in the mean-field case do we have $\beta_{\mbox {\scriptsize eff}} \equiv \beta_0$. The effective transmission rates  obtained directly from the simulations by  the method of  Section 3.3  were  $\beta_{\mbox {\scriptsize eff}} = 1.86$ for $p_{\mbox{\scriptsize SW}}= 0.2$ and $\beta_{\mbox {\scriptsize eff}} = 3.93$  in the homogeneous mixed case, in very good  agreement with the respective $\beta$'s calculated from Eq. (\ref{R0_SEIR}).

\subsection{Comparison between childhood diseases}

\begin{figure} [ht]
\centerline{\epsfxsize=3.3in\epsfbox{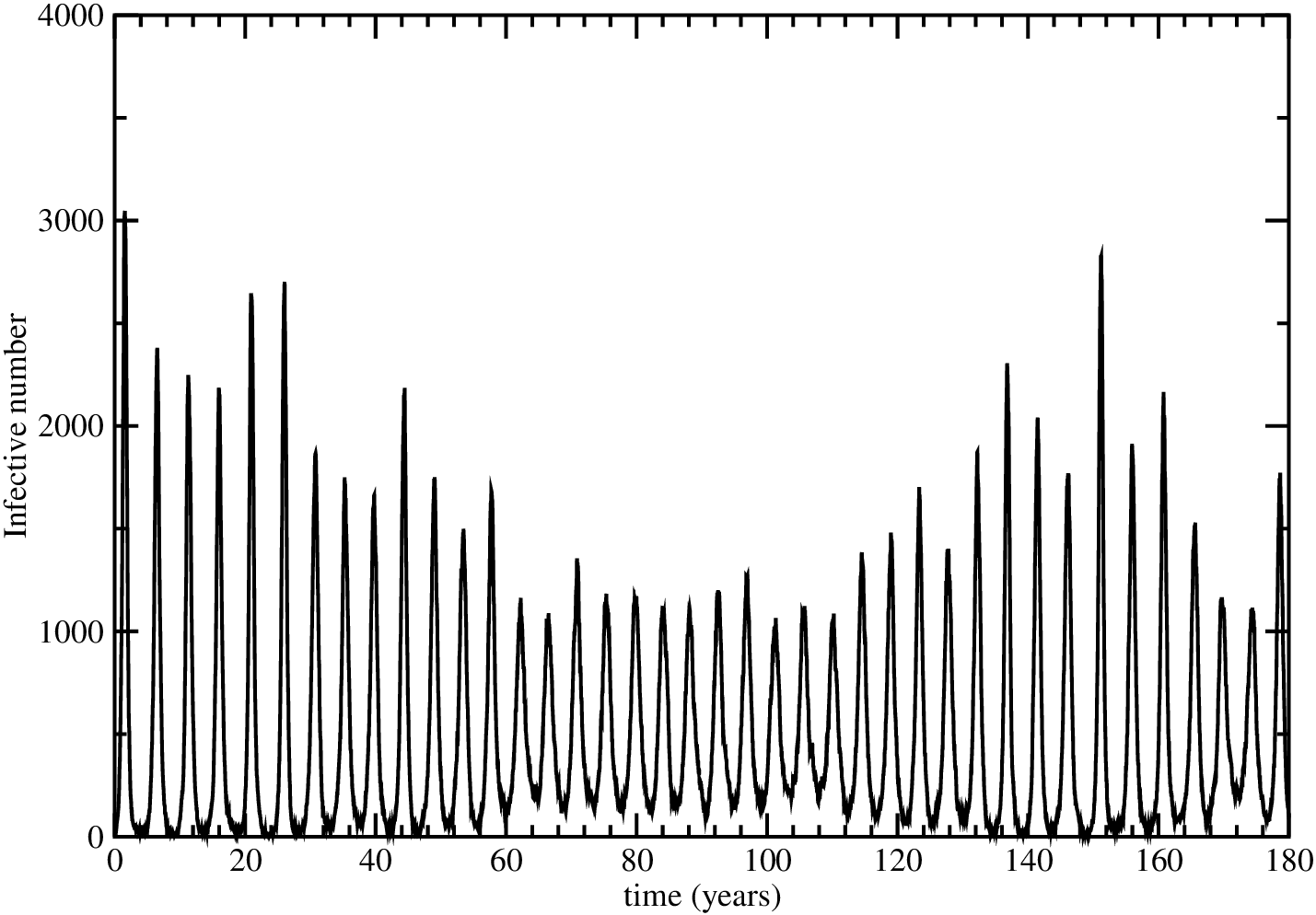}}
\caption{\small{Time series of the number of infectives for rubella, which  has longer latency and infectious periods than measles and also lower infectiousness. Simulation in a $1000 \times 1000$ lattice with $p_{\mbox{\scriptsize SW}}=0.2$. The birth rate  $\mu = 1/61$ yr$^{-1}$ is the same as in the  measles simulations, but now  $\tau_{lat} = 12$ days, $\tau_{inf} = 12$ days and $\beta_0 = 1.32$ day$^{-1}$.}} \label{rubella}
\end{figure}

\begin{figure} [ht]
\centerline{\epsfxsize=3.4in\epsfbox{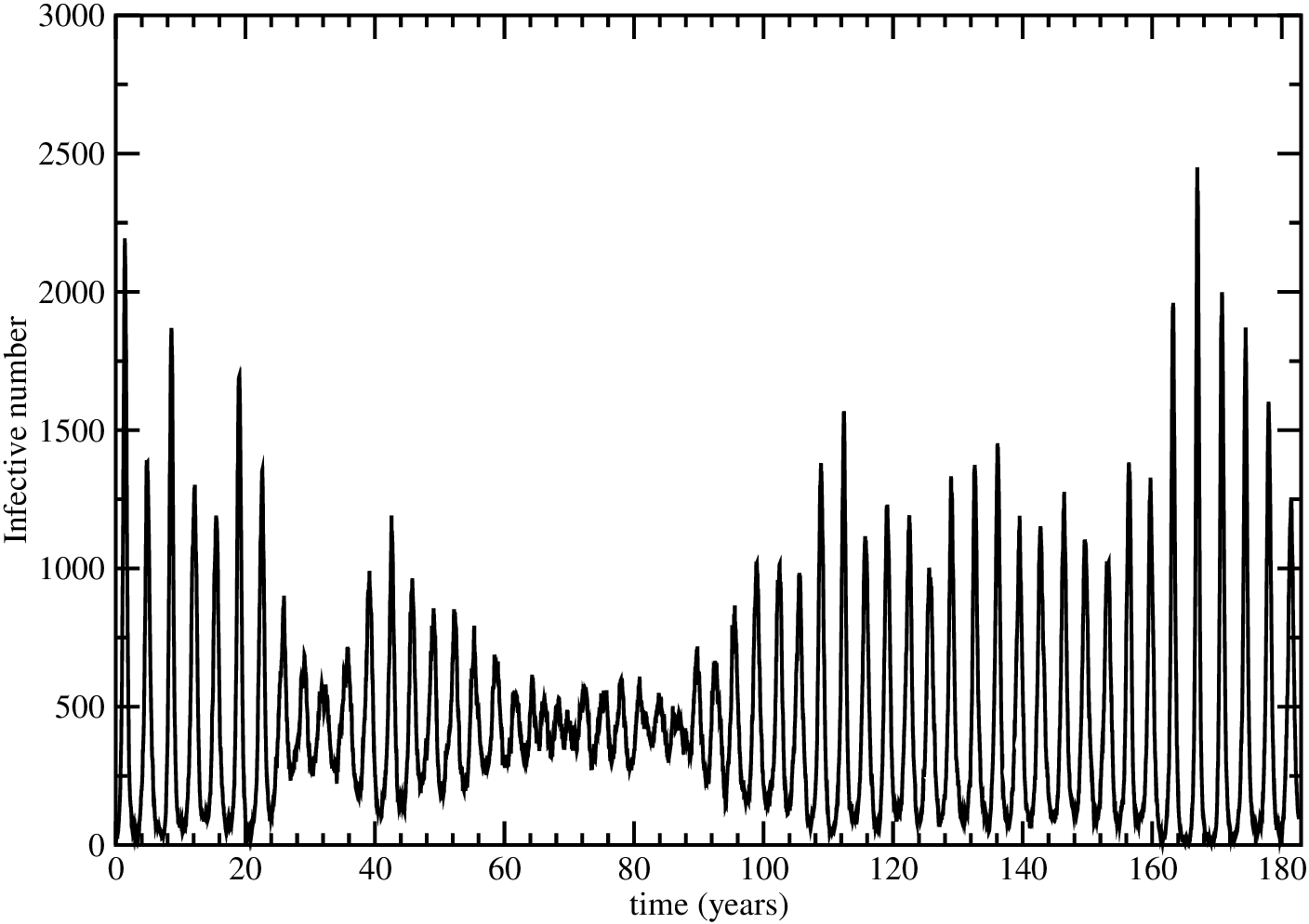}}
\caption{\small{Time series of the number of infectives for etiological parameters corresponding to chickenpox. Demographic parameters and small-world parameter are the same as for rubella and measles;  $\tau_{lat} = 10$ days, $\tau_{inf} = 10$ days and $\beta_0 = 2.4$ day$^{-1}$.}} \label{chicken}
\end{figure}

In this section  we show  that the ability of the PCA  to describe sustained oscillations is not restricted to measles but  extends to  other  childhood diseases conferring life-long immunity. In Figs. \ref{rubella} and \ref{chicken} we show long-term runs for etiological  parameters corresponding to rubella and chickenpox, respectively.
We kept the small-world parameter fixed at  $p_{\mbox{\scriptsize SW}}= 0.2$, since childhood diseases must share a common contact structure. The PCA is nevertheless  capable of discriminating between these different diseases, in terms of period,  average age at infection and basic reproductive rate.

The periods estimated from the Fourier transform of the time series in Figs. \ref{rubella} and \ref{chicken} were $T \approx 4.4$ years for rubella and $T \approx 3.4$ for chickenpox;  the average age at infection was $A \approx 6.2$ for rubella and $A \approx 4.7$ for chickenpox. Like for measles, $A$ is considerably  underestimated by the PCA,   most studies in developed countries  giving a lower bound for $A$ of about  9 years for rubella and 6 years for chickenpox. Still, the relative values of $A$ for the three diseases agree with the data \cite{AM}. The basic reproductive rate computed from the inverse fraction of susceptibles is  about $10$ for rubella and $13$ for chickenpox. The simulations for measles gave $R_0 \approx 15$. Again, the values for measles and chickenpox are very good while the reproductive rate of rubella is above that reported. 

Based on a purely qualitative evaluation of the mode of transmission we can rank common   childhood diseases in roughly two classes of infectiousness. Measles and chickenpox are transmitted through aerosol droplets and therefore the most infectious.  In the second group we have  mumps and rubella which require direct contact with droplets generated by sneezing and coughing. In our  simulations, this hierarchy of infectiousness was strictly  respected, with \break  $\beta_0^{\mbox {\scriptsize measles}} > \beta_0^{\mbox {\scriptsize chickenpox}} > \beta_0^{\mbox {\scriptsize rubella}}$. The numerical values used in the simulations give a quantitative estimate of  infectiousness:
\begin{equation}
{\beta_0^{\mbox {\scriptsize measles}} \over \beta_0^{\mbox {\scriptsize rubella}}} = 3.7 , \ \ \ {\beta_0^{\mbox {\scriptsize measles}} \over \beta_0^{\mbox {\scriptsize chickenpox}}} = 1.9, \nonumber
\end{equation}
and  the respective  effective transmission rates obtained  for a fraction of \break 20 \% of  long-range contacts give:
 \begin{equation}
{\beta_{\mbox {\scriptsize eff}}^{\mbox {\scriptsize measles}} \over \beta_{\mbox {\scriptsize eff}}^{\mbox {\scriptsize rubella}}} = 2.5, \ \ \ {\beta_{\mbox {\scriptsize eff}}^{\mbox {\scriptsize measles}} \over \beta_{\mbox {\scriptsize eff}}^{\mbox {\scriptsize chickenpox}}} = 1.7. \nonumber
\end{equation}

Our results suggest that measles is  about three to four  times more infectious then  rubella  and about twice as infectious as chickenpox. Given the crudeness of these estimations we can consider that the ratios  agree quite well  with those reported  by Keeling \& Grenfell \cite{KG2000}: 3.4 and 2.4 respectively for the transmissibility ratios and 2.5 and 1.4 for the effective transmission rate.\footnote{Although their study focus on mumps instead of rubella, almost surely  the two diseases have very similar values of infectiousness.} In another study \cite{KRG} a value of 3.8  for the ratio of $\beta_0$'s between measles and rubella was obtained by  fitting the output of a seasonally forced SIR model to the data.

\vspace{8mm}

\section{Discussion}

The results in this paper show  that both the short and  long-term dynamics of  childhood  diseases conferring life-long immunity can be described by taking into account a  dynamic small-world  network of contacts. Moreover, the simulations make clear  that the ability of recurrent epidemics to survive for a large number of cycles depends strongly on the level of heterogeneity in contact structure. The fraction of runs that survive up to a maximum ascribed time  has a non-trivial dependence on the fraction of long-range contacts displaying, at low enough birth rates, a maximum at finite fraction of shortcuts.  On the other hand, the persistence varies with the population size, and this gives the rationale on which to base a network-dependent  Critical Community Size.  The oscillations get more synchronized and their amplitude is strongly enhanced when $p_{\mbox{\scriptsize SW}}$ is decreased below 1, {\em i.e.} as the weight of local contacts is increased. The amplitude cannot keep growing  because the troughs between epidemic surges will get so deep, the number of surviving infectives so close to zero, that a stochastic fluctuation  will eventually drive the epidemic extinct.  This  explains why an increase in amplitude is correlated with a decrease in persistence.

The small-world network allows us to go step by step from a dynamical system with a large number of degrees of freedom --  which for $p_{\mbox{\scriptsize SW}} = 0$ are the $3N$ possible states of the lattice -- to the case  where the evolution  can be captured by only a couple of  ordinary differential equations. How this this contraction of phase space happens is an important but difficult question that deserves further investigation. We do know, however, the properties of the flow of the  low dimensional deterministic system, namely that for the SIR ODE's  the only attractor  is a fixed point $(s_0, i_0)$ and that there is a  saddle point at $(1,0)$ with its stable manifold along the $i_0 = 0$ axis. The fraction of infectives in the endemic steady-state  is proportional to the birth rate,  $i_0 = \mu(R_0 - 1)/\beta$, therefore  as $\mu \rightarrow 0$, $i_0$ gets asymptotically close to the stable manifold.  Since the real part of the pair of complex conjugate eigenvalues vanishes at zero birth rates one observes the  critical slowing down of the dynamics at $\mu \rightarrow 0$. Small disturbances such as those provided by the stochastic drive will enable the system to  make   a large  excursion in phase space before returning again to the vicinity of $(s_0, i_0)$ where it is never allowed to settle.  

It is clear that the progressive introduction of  spatial degrees of freedom brings spatio-temporal coherence into the stochastic dynamics. The oscillations at small $p_{\mbox{\scriptsize SW}}$ display a strong  contribution from  higher-order harmonics indicating that the system visits orbits located  further away from the fixed point.  Also, both the amplitude and period are larger  than at larger  values of $p_{\mbox{\scriptsize SW}}$, the oscillations evolving on a  slower time  scale compared to the less structured ones  observed in the limit of  random mixing. Clustering  lowers the effective transmission rate and the average infective number but has the  collateral  effect of  deepening the troughs and rising the peaks of  oscillations. In the epidemic lows,  infectives and susceptibles keep close together and the epidemic almost dies out. But one shortcut that randomly links an infective to the middle of  a susceptible cluster will be  enough to cause a large  outbreak that may even  consume  every susceptible in that cluster. However some susceptibles will still  remain  in  small clusters scattered all over the lattice and, screened from infection, their numbers will steadily grow at the  (slow) pace dictated by the  birth rate. Eventually one or more of these clusters will reach a size large enough  that  there will be  a high   probability of a long-range infection event linking to an individual inside them.  When this happens the conditions are set for the cycle to  repeat itself. When the fraction of shortcuts is  high, susceptibles are steadily consumed at an intermediate rate and do not have time to aggregate into medium-size clusters.  On the average the infection does spread more effectively -- the average infective number is higher --   but the large outbreaks are suppressed and what we observe instead are  small, rapid fluctuations around an endemic state.


The simple  fact that the PCA implementation of SIR and SEIR dynamics leads to  sustained, fully developed oscillations as a consequence of heterogeneity in the  network of contacts is by far the most impressive difference between our results and the output of deterministic mean-field models. However there are quantitative differences arising in quantities that are  very  relevant for epidemiologists. The effective transmission rate  $\beta_{\mbox {\scriptsize eff}}$ is always smaller than the mean-field value $\beta_0$. In this respect our long-term simulations corroborate the  results obtained  from the simulation of a single epidemic outbreak \cite{KleGren}.  $R_0$  is also lower than in the limit of random mixing.  Moreover, in structured populations it is very important to  distinguish between measurements of $R_0$  obtained from long time series and those focusing  on a single outbreak because these two methods were  shown to present  the greater differences.

The SIR implementation can be used to access the qualitative impact of the network structure but to make the  predictions of the PCA  quantitative  SEIR dynamics  is required. Long time series, more than twice the length of the best available records can be easily obtained for realistic values of the model parameters and show  a  good agreement with observed values of the period and basic reproductive rate for measles and chickenpox. For rubella the agreement was only reasonable. In all the  three cases the average age  at infection is systematically  underestimated, the discrepancy  reaching 30 \% in the case of rubella. The values obtained could be improved by accounting for immunity by maternal antibodies but still with this respect the PCA   behaves  no better than  the  mean-field models. 

 Different childhood diseases evolve on social networks that are similarly organized, transmission occurring predominantly in 
schools and households with a few exceptions corresponding to the long-range contacts. The structure of the contact patterns is thus identical and so are the demographic parameters.  A realistic network  model of  childhood diseases  must be able to discriminate between different diseases, in terms of $T$, $A$ and $R_0$, purely as a result of the different latency and infectious periods plus the transmissibility $\beta_0$. The PCA  satisfies this requirement and yields   a reasonable estimate for the infectiousness of each  disease. 

\section{Conclusion}

The individual-based network model presented in this paper combines local structure with   casual, long-range links. The latter are  shortcuts through which the disease can propagate into regions where susceptibles predominate. Correlations that build up in the system due to network structure  cause  deviations from mean-field behaviour, but  in the relevant limits the mean-field results are  restored. Surprisingly, we found that, when the   long-term evolution is considered, even with a  small fraction of shortcuts the  mean-field relations between the average age at infection, the basic reproductive rate  and the  average number of susceptibles still hold. This consistency may explain why, although based on a unrealistic description of human contacts, deterministic models featuring homogeneous mixing remained for so long  the basic conceptual tool of theoretical epidemiology. Their flaws, particularly the inability  to describe recurrent epidemics, were exposed in the present work, in the  context of  childhood diseases. These are highly infectious diseases for which it is not unreasonable to assume that anyone who engages in any  basic form of social interaction is equally at risk. Thus, we have disclosed  the weaknesses of mean-field models in the case where they are certainly   the less severe.  Simulation of  individual-based stochastic models  becomes  imperative in order to capture the  dynamical complexity  of infections like HIV of Hepatitis that spread on networks characterized by an extreme heterogeneity, like   the network  of sexual partnerships or needle sharing by intravenous drug users.


\section*{Acknowledgments}

The author received a fellowship from Funda\c{c}\~{a}o para a Ci\^{e}ncia e Tecnologia (FCT), ref. SFRH/BPD/5715/2001. Financial support  by the FCT under project  POCTI/ESP/44511/2002 is also gratefully acknowledged.


%

\newpage

\newpage


\begin{thebibliography}{99}

\bibitem{AM}{Anderson and May (1991), Infectious Diseases of Humans, Oxford University Press, Oxford.}






\bibitem{ERBG}{Earn, David J. D. {\em et al.} 
(2000), A  Simple Model for Complex Dynamical Transitions in Epidemics, {\em Science} {\bf 287}, 667-670.}


\bibitem{GKB}{Greenman, Jon, Kamo, Masashi, and Boots, Mike (2004), External Forcing of Ecological and Epidemiological Systems: a Resonance Approach, {\em Physica} D {\bf 190}, 136-151.}

\bibitem{KRG}{Keeling, Matt J., Rohani, Pejman and Grenfell, Bryan T. (2001), Seasonally Forced Disease Dynamics Explored as Switching Between Attractors, {\em Physica} D, {\bf 148}, 317-335.}


\bibitem{OlsenShaff}{Olsen, L. F. and Schaffer, W. M. (1990), Chaos versus Noisy Periodicity: Alternative Hypothesis for Childhood Epidemics, {\em Science} {\bf 249}, 499-504}.


\bibitem{RohKeelGren}{Rohani, Pejman, Keeling, Matthew J. and Grenfell, Bryan T. (2002), The Interplay between Determinism and Stochasticity in Childhood Diseases, {\em The American Naturalist}, {\bf 159}, 469-481.}



\bibitem{Blasius}{Blasius, Bernd, Huppert, Amit and Stone, Lewi (1999), Complex Dynamics and Phase Synchronization in Spatially Extended Ecological Systems, {\em Nature}, {\bf 399}, 354-359.}

\bibitem{Bailey}{Bailey, Norman T. J. (1975), The Mathematical Theory of Infectious Diseases, Charles Griffin, London.}

\bibitem{LloydMay}{Lloyd, Alen L. and May, Robert M. (1996), Spatial Heterogeneity in Epidemic Models, {\em J. Theor. Biol.}, {\bf 179}, 1-11.}

\bibitem{BolkGren}{Bolker, Benjamin and Grenfell, Bryan (1995), Space, Persistence and Dynamics of Measles Epidemics, {\em Phil. Trans. R. Soc. Lond.} B, {\bf 348}, 309-320.}



\bibitem{KeelRohGren}{Keeling, Matt J., Rohani, Pejman and Grenfell, Bryan T.  (2000), Seasonally Forced Disease Dynamics Explored as Switching between Attractors, {\em Physica D}, {\bf 148}, 317-335.}


\bibitem{WS}{Watts, D. J.  and Strogatz, S. H. (1998), Collective Dynamics of 'Small-world' Networks, {\em Nature}, {\bf 393}, 440-442} 

\bibitem{Watts}{Watts, D. J., Small Worlds (1999), {Princeton University Press, Princeton, NJ}. }

\bibitem{Moore}{Moore, Christopher and Newman, M. E. J. (2000), Epidemics and Percolation in Small-world Networks, {\em Physical Review} E, {\bf 61}, 5678-5682.}

\bibitem{Newman}{Newman, M. J., Jensen, I. and Ziff, R. M. (2002), Percolation and Epidemics in a Two-dimensional Small World, {\em Physical Review} E, {\bf 65}, 021904.}


\bibitem{BS}{Boots, Michael and Sasaki, Akira  (1999), ``Small worlds'' and the evolution of virulence: infection occurs locally and at a distance, {\em Proc. R. Soc. Lond. }B, {\bf 266}, 1933-1938.}

\bibitem{KleGren}{Kleczkowski, A. and Grenfell, Bryan T., (1999), Mean-field-type equations for spread of epidemics: the ``small world'' model, {\em Physica A} {\bf 274}, 355-360.}


\bibitem{Keeling99}{Keeling, M. J. (1999), The effect of local spatial structure on epidemiological invasions, {\em Proc. R. Soc. Lond. }B, {\bf 266}, 859-867.}

\bibitem{Johan1}{Johansen, Anders (1996), A Simple Model of Recurrent Epidemics, {\em J. Theor. Biol.}, {\bf 178}, 45-51.}

\bibitem{Johan2}{Johansen, Anders (1994), Spatio-temporal Self-organization in a Model of Disease Spreading, {\em Physica} D {\bf 78}, 186-193.}

\bibitem{Kuperman}{Kuperman, Marcelo and Abramson, Guillermo, Small World Effect in an Epidemiological Model, {\em Phys. Rev. Lett.}, {\bf 86}, 2909-2911.}


\bibitem{HeStone}{He, David and Stone, Lewi (2003), Spatio-temporal synchronization of recurrent epidemics,  {\em Proc. R. Soc. Lond. }B, {\bf 270}, 1519-1526.}


\bibitem{VerdascaJTB}{Verdasca, J. {\em et al.} (2005), Recurrent epidemics in small world networks, {\em J. Theor. Biol.}, {\bf 233}, 553-561}.



\bibitem{Lloyd01}{Lloyd, Alun L. (2001), Realistic Distribution of Infectious Periods in Epidemic Models: Changing Patterns of Persistence and Dynamics, {\em Theor. Pop. Biol.}, {\bf 60}, 59-71.}

\bibitem{SARSRiley}{Riley, S. R. et al., Transmission Dynamics of the Etiological Agent of SARS in Hong Kong: Impact of Public Health Interventions, {\em Science} {\bf 300}, 1961-1966.}

\bibitem{SARSLip}{Lipsitch, M. et al., Transmission Dynamics and Control of Severe Acute Respiratory Syndrome, {\em Science} {\bf 300}, 1966-1970}


\bibitem{Diekmann}{Diekmann O. and Heesterbeek, J. A. P. (2000), Mathematical Epidemiology of Infectious Diseases, Wiley, New York.} 

\bibitem{KG2000}{Keeling, Matt J. and Grenfell, Bryan T. (2000), Individual-based Perspectives on $R_0$, {\em J. theor. Biol.}, {\bf 203}, 51-61.}














\end{thebibliography}
\end{document}